# Multi-plane denoising diffusion-based dimensionality expansion for 2D-to-3D reconstruction of microstructures with harmonized sampling

Kang-Hyun Lee[1], and Gun Jin Yun[1,2,*]

[1]*Department of Aerospace Engineering, Seoul National University, Gwanak-gu Gwanak-ro 1 Seoul 08826, South Korea*
[2]*Institute of Advanced Aerospace Technology, Seoul National University, Gwanak-gu Gwanak-ro 1, Seoul 08826, South Korea*

**Abstract**

Acquiring reliable microstructure datasets is a pivotal step toward the systematic design of materials with the aid of integrated computational materials engineering (ICME) approaches. However, obtaining three-dimensional (3D) microstructure datasets is often challenging due to high experimental costs or technical limitations, while acquiring two-dimensional (2D) micrographs is comparatively easier. To deal with this issue, this study proposes a novel framework for 2D-to-3D reconstruction of microstructures called 'Micro3Diff' using diffusion-based generative models (DGMs). Specifically, this approach solely requires pre-trained DGMs for the generation of 2D samples, and dimensionality expansion (2D-to-3D) takes place only during the generation process (i.e., reverse diffusion process). The proposed framework incorporates a new concept referred to as 'multi-plane denoising diffusion', which transforms noisy samples (i.e., latent variables) from different planes into the data structure while maintaining spatial connectivity in 3D space. Furthermore, a harmonized sampling process is developed to address possible deviations from the reverse Markov chain of DGMs during the dimensionality expansion. Combined, we demonstrate the feasibility of Micro3Diff in reconstructing 3D samples with connected slices that maintain morphologically equivalence to the original 2D images. To validate the performance of Micro3Diff, various types of microstructures (synthetic and experimentally observed) are reconstructed, and the quality of the generated samples is assessed both qualitatively and quantitatively. The successful reconstruction outcomes inspire the potential utilization of Micro3Diff in upcoming ICME applications while achieving a breakthrough in comprehending and manipulating the latent space of DGMs

**Keywords:** Denoising diffusion; Microstructure reconstruction; Dimensionality expansion; 2D-to-3D reconstruction; Diffusion-based generative models

## 1. Introduction

The properties and physical behavior of a material are profoundly influenced by its microstructure, which encompasses the topology, distribution, and physical characteristics of the constituent phases. By leveraging computational mechanics [1, 2] and integrated computational materials engineering (ICME) approaches [3, 4], the relationship between the microstructure and properties can be systemically investigated for exploring the design space of materials [5-8]. For precise modeling and understanding of complex material behavior in terms of microstructure-property linkage, conducting three-dimensional (3D) analysis and

---

characterization of microstructures are imperative. Furthermore, the accuracy of computational analysis for investigating material properties relies on the quality of accessible 3D microstructural datasets. In order to acquire precise 3D information about microstructures, micro-computed tomography (micro-CT) has demonstrated its effectiveness as a tool allowing for the visualization of a material's internal structure and potential defects in a 3D domain. [9-12]. Moreover, the use of micro-CT-based characterization of microstructures has been regarded as a reliable approach to understand material behavior with computational analysis, such as finite element analysis (FEA), particularly at the representative volume element (RVE) scale [9, 12-16].

However, the difficulty lies in acquiring an extensive database of microstructures with experimental analysis, primarily due to time and cost constraints. Furthermore, obtaining a 3D microstructural dataset is more challenging, as it requires serial sectioning views, in contrast to the relatively straightforward capturing independent two-dimensional (2D) micrographs. To address this challenge, numerous microstructure characterization and reconstruction (MCR) approaches employing a diverse range of microstructural descriptors have been proposed [13, 17-24]. In general, these descriptor-based MCR methods obtain morphologically equivalent samples by iterative optimization of the discrepancy between the target and the current descriptors. Various kinds of descriptors, including simple volume fraction as well as high-dimensional correlation functions such as n-point correlations [17, 25] or lineal path functions [26], can be employed to quantify the morphology of the microstructure. Once specific descriptors are chosen, the optimization problem can be addressed using stochastic reconstruction techniques like the Yeong-Torquato algorithm [24]. Moreover, it is possible to utilize differentiable descriptors to reconstruct equivalent 3D microstructural samples through the application of a gradient-based optimizer [17, 27]. In particular, Seibert et al. [20] proposed a 2D-to-3D reconstruction method to obtain realistic microstructure samples using differentiable microstructural descriptors and optimization algorithms provided by the open-access MCRpy package [19].

Although MCR methods with microstructural descriptors have been demonstrated to be effective for microstructure reconstruction, a major concern of this approach is that it requires specific descriptors to be optimized. Thus, for achieving favorable reconstruction outcomes, the selection of appropriate descriptors is crucial considering the specific morphological attributes of the targeted microstructure. Meanwhile, there has been extensive research on generative models, such as variational autoencoders (VAEs) [28-31] and generative adversarial networks (GANs) [32-34], for reconstructing microstructures by learning the underlying distribution of data. For instance, Kim et al. [28] proposed a VAE-based framework for reconstruction of 2D microstructure samples from a deep-learned continuous microstructure space. They also demonstrated that inverse design is possible by establishing connections between the features in the latent space of VAE and the mechanical properties of materials. Fokina et al. [49] utilized the style-based GAN architecture for reconstruction of 2D microstructural samples that are close to the original samples in terms of area density and Euler characteristics distributions. In addition, 2D-to-3D microstructure reconstruction with generative models has been gaining significant attention recently [34-36]. One of the remarkable works is the novel GAN architecture called SliceGAN, proposed by Kench and Cooper [34], which synthesizes 3D microstructure datasets from a single representative 2D image. They demonstrated their GAN-based model can effectively reconstruct 3D microstructures of various types of materials, including polycrystalline metals, ceramics, and battery electrodes

On the other hand, VAEs suffer from a notable drawback, as the generated samples tend to be distorted and blurred [37, 38], while GANs are susceptible to the issue of mode collapse and unstable training due to the adversarial loss function [39, 40]. In light of these concerns, the diffusion-based generative models (DGMs) [41-43] are currently gaining significant attention as a promising state-of-the-art generative model. In particular, there are two popular formulations of DGMs which are called the denoising diffusion probabilistic models (DDPMs) and the score-based generative models (SGMs). According to the formulation of DDPMs [43], a DGM can be perceived as a model of reverse Markovian chain, intended for progressive denoising a data structure, while SGMs [42] view DGMs as models for estimating gradients of data distribution with denoising score matching [44]. The DGMs can also be generalized to the problem of solving reverse stochastic differential equations (SDEs) in order to transform the noise distribution into the data distribution [41, 45]. Due to the transformation of generation problem into the progressive reverse diffusion (i.e., denoising) process, DGMs have shown superior performance compared to GANs in generating high-quality images [46], and they are not prone to mode collapse or unstable training which are commonly observed in GANs.

However, the possibility of using DGMs for 2D-to-3D reconstruction of microstructures has not been explored, according to the best knowledge of the authors. While some studies have suggested the use of DGMs for reconstructing 2D microstructure samples [5, 47, 48], the problem of dimensional expansion (2D-to-3D) has not been explored extensively due to the lack of knowledge in the latent space of DGMs. In order to bridge this gap, a novel dimensionality-expansion framework for 2D-to-3D reconstruction of microstructures based on denoising diffusion, called 'Micro3Diff', is proposed in this study. The proposed method does not require any 3D datasets, and it only requires a DGM trained for the generation of 2D images (2D-DGM). The feasibility of this approach lies in the fact that Micro3Diff targets only the manipulation of the reverse diffusion process for dimensionality expansion. In addition, this greatly facilitates the application of this framework as the commonly used 2D-DGM can be easily incorporated. The structure of this article with the key contributions of this work are summarized as follows.

(1) In the 'Preliminaries' section, the formulations of DGMs are introduced along with the SDE-based generalization to provide an understanding of the basic theory of DGMs. The concept of transforming noise distribution into data distribution through progressive denoising is then employed to support the assumptions required for Micro3Diff, which is introduced in the next section.

(2) In the following section, the detailed procedure for 2D-to-3D reconstruction using Micro3Diff is introduced, based on the novel multi-plane denoising diffusion-based approach. Furthermore, to mitigate potential discrepancies arising from multi-plane denoising, the technique of harmonized sampling is introduced to enhance the quality of generated samples. By combining these approaches, this section demonstrates how it becomes feasible to simultaneously transform the noise distribution in multiple planes into the data distribution while preserving connectivity and continuity in 3D space.

(3) To validate the proposed methodology, various types of microstructures including spherical inclusions, polycrystalline grains, battery electrodes, and carbonates, sourced from both synthetic and experimental data, are considered to be reconstructed using the proposed Micro3Diff. The generated results are subsequently validated by comparing spatial correlation functions to assess the similarity between the generated 3D samples and the original data.

## 2. Preliminaries

In this section, the formulations of DGMs are introduced, encompassing the two predominant approaches (SGMs and DDPMs), for the development of DGM-based 2D-to-3D reconstruction of microstructures (section 3).

### 2.1 Generalization of DGMs

Over the past few years, there have been several formulations and variations of DGMs [41-43, 49-54] proposed, yet they all share two important common features: 1. the pre-defined progressive noising process for transforming data distribution into a prior noise distribution, and 2. the incorporation of a model that learns to denoise a sample to transform the noise distribution towards a desired conditional/unconditional data distribution. The formal is also called the forward diffusion process and the latter one is called the reverse diffusion process [45, 54]. Among the different formulations, the one that employs the SDEs can be regarded as a generalized form of DGMs. A forward diffusion process for perturbing (i.e., noising) data $\mathbf{x}$ can be defined in a form of Itô SDE [55] with a continuous time variable $t \in [0, T]$ as

$$\mathrm{d}\mathbf{x} = \mathbf{f}(\mathbf{x}, t) + \sigma(t)\mathrm{d}\boldsymbol{\omega} \tag{1}$$

where $\boldsymbol{\omega}$ is a standard Wiener process, which can also be thought of as Gaussian noise or Brownian motion [41, 56], $\mathbf{f}(\cdot, t): \mathbb{R}^n \to \mathbb{R}^n$ and $\sigma(t): \mathbb{R} \to \mathbb{R}$ represent drift and diffusion coefficients of SDE at time step $t$, respectively. To perturb data with the SDE, the drift coefficient can be defined to nullify the data while the diffusion coefficient is adopted to progressively add noise to the data. In other words, the diffusion process can be modeled with different choices of $\mathbf{f}(\cdot, t)$ and $\sigma(t)$ while $\mathrm{d}\boldsymbol{\omega}$ represents the stochastic component. Considering this aspect, the two popular formulations of DGMs, which are SGMs [42] and DDPMs [43], can also be interpreted as discretized variations of the SDE with different drift and diffusion coefficients.

### 2.2 Forward diffusion process

As stated above, the diffusion processes in the various formulations of DGMs (e.g., SGMs and DDPMs) can be generalized using the concept of SDE. For instance, SGMs employ the concept of the score function (i.e., the gradient of the log probability density) and the denoising score matching with Langevin dynamics [44] for transforming random noise to data with high probability density. According to the perturbation process with a specified noise distribution proposed by Song et al. [42], the SDE for SGMs can be written as

$$\mathrm{d}\mathbf{x} = \sqrt{\frac{d[g(t)^2]}{dt}}\mathrm{d}\mathbf{w} \tag{2}$$

where $g\left(\frac{t}{T}\right)$ converges to $g_t$ (i.e., the noise level at time $t$) as $T \to \infty$. Considering a discrete sequence of samples (i.e., $\mathbf{x}_0, \mathbf{x}_1, \ldots, \mathbf{x}_T$), the update rule for obtaining the noised sample ($\mathbf{x}_{t+1}$) from the sample at the previous time step ($\mathbf{x}_t$) can be defined as follows with the noise component $\boldsymbol{\varepsilon}_t \sim N(\mathbf{0}, \mathbf{I})$.

$$\mathbf{x}_{t+1} = \mathbf{x}_t + \sqrt{g_{t+1}^2 - g_t^2}\,\boldsymbol{\varepsilon}_t \tag{3}$$

On the other hand, DDPMs utilize the concept of Markov chain to define the forward/reverse diffusion process, along with the variational lower bound (VLB) for training the models to reconstruct an original data structure from the prior noise distribution. Similar yet different, the SDE for the forward diffusion process of DDPMs can be expressed as

$$d\mathbf{x} = -\frac{1}{2}\beta(t)\mathbf{x}dt + \sqrt{\beta(t)}d\mathbf{w} \qquad (4)$$

where $\beta$ is the noise scheduling parameter and $\beta\left(\frac{t}{T}\right)$ converges to $T\beta_t$ as $T \to \infty$. According to this definition and the Stirling's approximation, the following update rule can be derived.

$$\mathbf{x}_{t+1} = \sqrt{1-\beta_t}\mathbf{x}_t + \sqrt{\beta_t}\boldsymbol{\varepsilon}_t \qquad (5)$$

The schematic of the forward diffusion process of DGMs using the introduced formulations is shown in Figure 1. It is again worth noting that the primary objective of defining the forward diffusion process in DGMs is to transform the data distribution into the prior noise distribution, even though the drift and diffusion coefficients may vary.

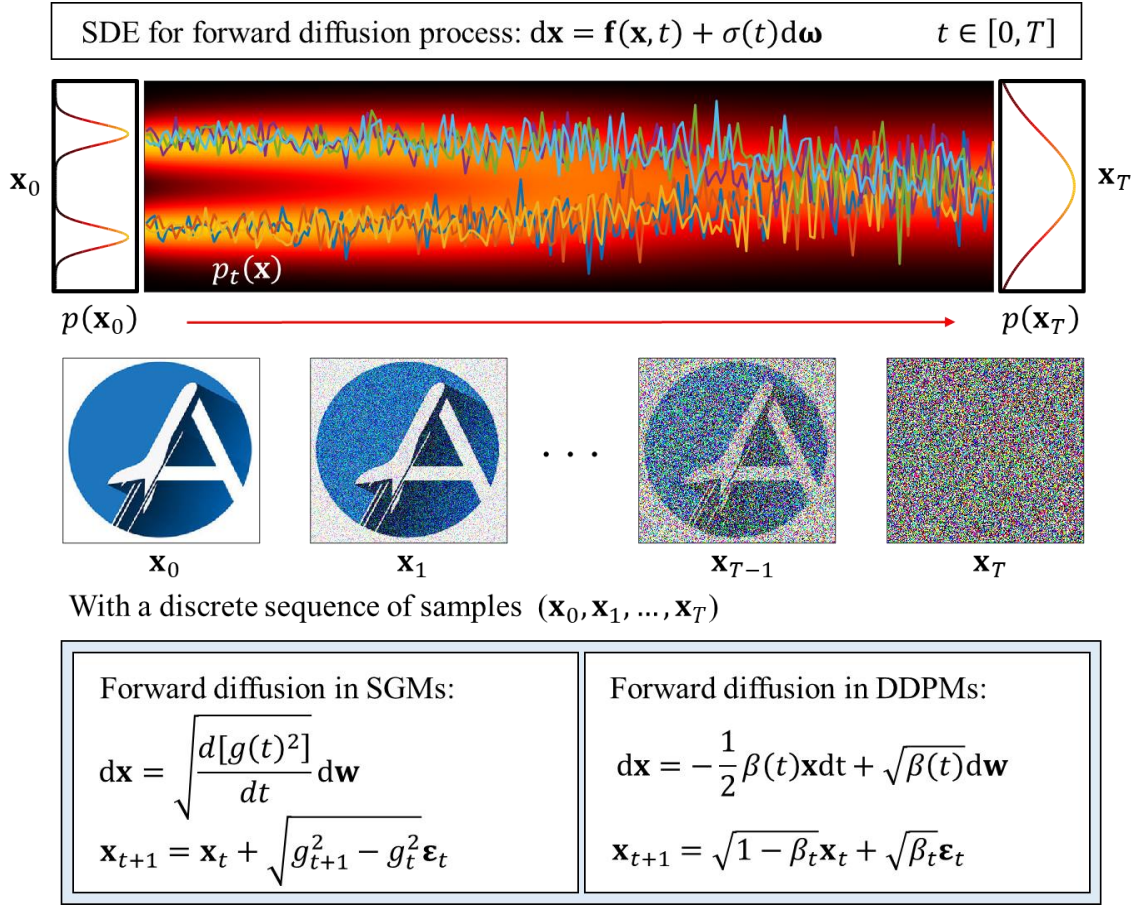

**Figure 1.** Schematic of the forward diffusion process (i.e., noising process) with the formulations based on SDE, SGM and DDPM.

## 2.3 Reverse diffusion process

Based on the forward diffusion SDE (Eq. (1)), the SDE for reverse diffusion of data can be derived as

$$d\mathbf{x} = [\mathbf{f}(\mathbf{x}, t) - \sigma(t)^2 \nabla_\mathbf{x} \log p_t(\mathbf{x})]dt + \sigma(t)d\hat{\boldsymbol{\omega}} \qquad (6)$$

where $\hat{\boldsymbol{\omega}}$ is a standard Wiener process when the time is reversed. In an intuitive sense, the SDE in reversed time (Eq. (6)) implies that it is able to reconstruct data by starting with the noise distribution. It is worth noting that the only unknown in Eq. (6) is the value of score function, which is denoted as $\nabla_\mathbf{x} \log p_t(\mathbf{x})$. In other words, if we can obtain the value of $\nabla_\mathbf{x} \log p_t(\mathbf{x})$, it is able to generate as many samples as desired. In this regard, the following loss function of SGMs for denoising score matching [42] can be optimized:

$$L_{\text{SGM}} := \mathbb{E}_{\mathbf{x}_t \sim p(\mathbf{x}_t)} \left[ \lambda(t) g_t^2 \| \nabla_{\mathbf{x}_t} \log p(\mathbf{x}_t) - \mathbf{s}_\theta(\mathbf{x}_t, t) \|^2 \right] \qquad (7)$$

where $\lambda(t)$ is a positive weighting function at time $t$ and $\mathbf{s}_\theta(\mathbf{x}_t, t)$ is a model trained to estimate the value of the score function within the discrete time interval. In addition, this can be considered equivalent to solving the reverse-time SDE.

On the other hand, the loss function of DDPMs originates from the theory of variational inference [43, 57]. This is the reason why a DDPM is also called a hierarchical Markovian VAE [45]. Based on the Eq. (5), a process of forward diffusion can be defined in the form of conditional Gaussian distribution as follows.

$$p(\mathbf{x}_t | \mathbf{x}_{t-1}) = N\left(\mathbf{x}_t; \sqrt{1 - \beta_t}\mathbf{x}_{t-1}, \beta_t \mathbf{I}\right) \qquad (8)$$

Based on this forward process, a neural network model for the reverse process can be defined as

$$p_\theta(\mathbf{x}_{t-1} | \mathbf{x}_t) = N\left(\mathbf{x}_{t-1}; \boldsymbol{\mu}_\theta(\mathbf{x}_t, t), \boldsymbol{\Sigma}_\theta(\mathbf{x}_t, t)\right) \qquad (9)$$

where $\boldsymbol{\mu}_\theta$ and $\boldsymbol{\Sigma}_\theta$ are the predicted mean function and the covariance, respectively. To estimate the original data distribution $p(\mathbf{x}_0)$ with the forward/reverse process through the time steps, the Kullback-Leibler divergence ($D_{KL}$) between the joint distributions $p_\theta(\mathbf{x}_0, \mathbf{x}_1, \ldots, \mathbf{x}_T)$ and $p(\mathbf{x}_0, \mathbf{x}_1, \ldots, \mathbf{x}_T)$ need to be minimized. This is achieved by minimizing the VLE of the negative log-likelihood as:

$$L_{\text{vlb}} := L_0 + L_1 + \cdots + L_{T-1} + L_T \qquad (10)$$

$$L_0 := -\log p_\theta(\mathbf{x}_0 | \mathbf{x}_1) \qquad (11)$$

$$L_{t-1} := D_{KL}\left(p(\mathbf{x}_{t-1} | \mathbf{x}_t, \mathbf{x}_0) \| p_\theta(\mathbf{x}_{t-1} | \mathbf{x}_t)\right) \qquad (12)$$

$$L_T := D_{KL}\left(q(\mathbf{x}_T | \mathbf{x}_0) \| p(\mathbf{x}_T)\right) \qquad (13)$$

In particular, $L_{t-1}$ can be rewritten as the following expectation of $\ell_2$ loss as

$$L_{t-1} = \mathbb{E}_p[\lambda(t)\|\boldsymbol{\mu}_t(\mathbf{x}_t, \mathbf{x}_0) - \boldsymbol{\mu}_\theta(\mathbf{x}_t, t)\|^2] \tag{14}$$

where $\boldsymbol{\mu}_\theta(\mathbf{x}_t, t)$ is the model trained to estimate the mean function and $\boldsymbol{\mu}_t(\mathbf{x}_t, \mathbf{x}_0)$ is the mean function of noised sample at $t$ according to the pre-defined forward diffusion process (Eq. (5)). In particular, Ho et al. [43] reparametrized Eq. (14) with the model $\boldsymbol{\varepsilon}_\theta(\mathbf{x}_t, t)$ for predicting the noise component at $t$ as follows.

$$\mathbb{E}_{t\sim[1,T],\mathbf{x}_0\sim p(\mathbf{x}_0),\boldsymbol{\varepsilon}\sim N(\mathbf{0},\mathbf{I})}[\lambda(t)\|\boldsymbol{\varepsilon} - \boldsymbol{\varepsilon}_\theta(\mathbf{x}_t, t)\|] \tag{15}$$

To show the linkage between DDPMs and SGMs, Eq. (7) can be rewritten according to Eq. (2) and (3) as

$$\begin{aligned}
L_{\text{SGM}} &= \mathbb{E}_{t\sim[1,T],\mathbf{x}_0\sim p(\mathbf{x}_0),\mathbf{x}_t\sim p(\mathbf{x}_t)}\left[\lambda(t)g_t^2\|\nabla_{\mathbf{x}_t}\log p(\mathbf{x}_t|\mathbf{x}_0) - \mathbf{s}_\theta(\mathbf{x}_t, t)\|^2\right] \\
&= \mathbb{E}_{t\sim[1,T],\mathbf{x}_0\sim p(\mathbf{x}_0),\mathbf{x}_t\sim p(\mathbf{x}_t)}\left[\lambda(t)\left\|-\frac{\mathbf{x}_t - \mathbf{x}_0}{g_t} - g_t\mathbf{s}_\theta(\mathbf{x}_t, t)\right\|^2\right] \\
&= \mathbb{E}_{t\sim[1,T],\mathbf{x}_0\sim p(\mathbf{x}_0),\boldsymbol{\varepsilon}\sim N(\mathbf{0},\mathbf{I})}[\lambda(t)\|\boldsymbol{\varepsilon} + g_t\mathbf{s}_\theta(\mathbf{x}_t, t)\|^2]
\end{aligned} \tag{16}$$

which shows that DDPMs are equivalent to SGMs if $\boldsymbol{\varepsilon}_\theta(\mathbf{x}_t, t) = -g_t\mathbf{s}_\theta(\mathbf{x}_t, t)$.

### 3. Denoising diffusion-based 2D-to-3D reconstruction with harmonized sampling
3.1 Equation of sampling

In the previous section, the equivalence of the different formulations of DGMs is introduced. Among the different formulations, the formulation of DDPMs is adopted in this study to build DGMs for the reconstruction of microstructures. Thus, Eq. (15) is utilized as the objective function for training the model $\boldsymbol{\varepsilon}_\theta(\mathbf{x}_t, t)$. After training the model, it is able to obtain the mean function of $\mathbf{x}_t$ in Eq. (9) with the prediction of $\boldsymbol{\varepsilon}$ using the following equation:

$$\boldsymbol{\mu}_\theta(\mathbf{x}_t, t) = \frac{1}{\sqrt{\alpha_t}}\left(\mathbf{x}_t - \frac{\beta_t}{\sqrt{1-\bar{\alpha}}}\boldsymbol{\varepsilon}_\theta(\mathbf{x}_t, t)\right) \tag{17}$$

where

$$\alpha_t = 1 - \beta_t \tag{18}$$

$$\bar{\alpha}_t = \prod_{s=0}^{t}\alpha_s \tag{19}$$

It is worth noting that $\boldsymbol{\Sigma}_\theta$ is assumed to be constant, which can be simply computed using the pre-defined noise schedule (i.e., $\boldsymbol{\Sigma}_\theta = \beta_t\mathbf{I}$), as learning only the mean function leads to better sample quality according to Ho et al. [43]. In addition, the variance can be also learned by incorporating a parameterized diagonal $\Sigma_\theta$ into the VLE (Eq. (10)) [58].

3.2 Reverse diffusion process for dimensionality expansion (2D-to-3D)

One of the most distinct characteristics of DGMs, compared to other generative models such as VAEs [14, 15] and GANs [16, 17], is the progressive and gradual diffusion process. As discussed in the section 2, DGMs incorporate the forward/reverse diffusion process to estimate the data distribution $p(\mathbf{x}_0)$ and generate samples from the known prior distribution $p(\mathbf{x}_T)$. For instance, if a DGM for the generation of 2D images (i.e., 2D-DGM) is prepared, it can generate novel 2D images start from Gaussian noise through the diffusion times steps and the pre-defined noise schedule [46, 51]. In addition, this process is equivalent to the gradual transformation of $p(\mathbf{x}_0)$ to $p(\mathbf{x}_T)$ in 2D pixel space. Drawing from this perspective, this study introduces a novel method for 2D-to-3D dimensionality expansion through the multiplane denoising diffusion process (Figure 2 and Figure 3). As shown in Figure 2, we can define three orthogonal planes (yz, xz, xy-planes) and use a trained 2D-DGM for generation of a sample at each plane. According to the original formulation of DDPMs, the generation process of each sample on each plane is independent of the others, resulting in the creation of three distinct samples that follow the data distribution $p(\mathbf{x}_0)$. However, if each generation process (i.e., reverse diffusion process) at each plane proceeds concurrently and is conditioned on the others, we could potentially achieve samples with connectivity and continuity ensured at the junctions of the planes. Thus, the problem now becomes developing a new model for estimating the conditional distributions at each time step during the reverse diffusion process as:

$$p_\theta\left(\mathbf{x}_{t-1}^{(*)} | \mathbf{x}_t^{(1)}, \mathbf{x}_t^{(2)}, \mathbf{x}_t^{(3)}\right) \tag{20}$$

where $\mathbf{x}_{t-1}^{(*)}$ denotes the denoised sample at a certain plane, and $\mathbf{x}_t^{(1)}, \mathbf{x}_t^{(2)}$ and $\mathbf{x}_t^{(3)}$ represent the samples before denoising at the three orthogonal planes. However, to train the model in Eq. (20), it is evident that a 3D dataset with samples at three orthogonal planes is required. In addition, the model requires additional channels or modules to impose conditional signals during the generation process, which necessitates modifying the model architecture, such as incorporating classifiers or using classifier-free guidance [46, 59]. Thus, this cannot be classified as 2D-to-3D reconstruction because both the dataset and the training process need to extend beyond the confines of 2D space

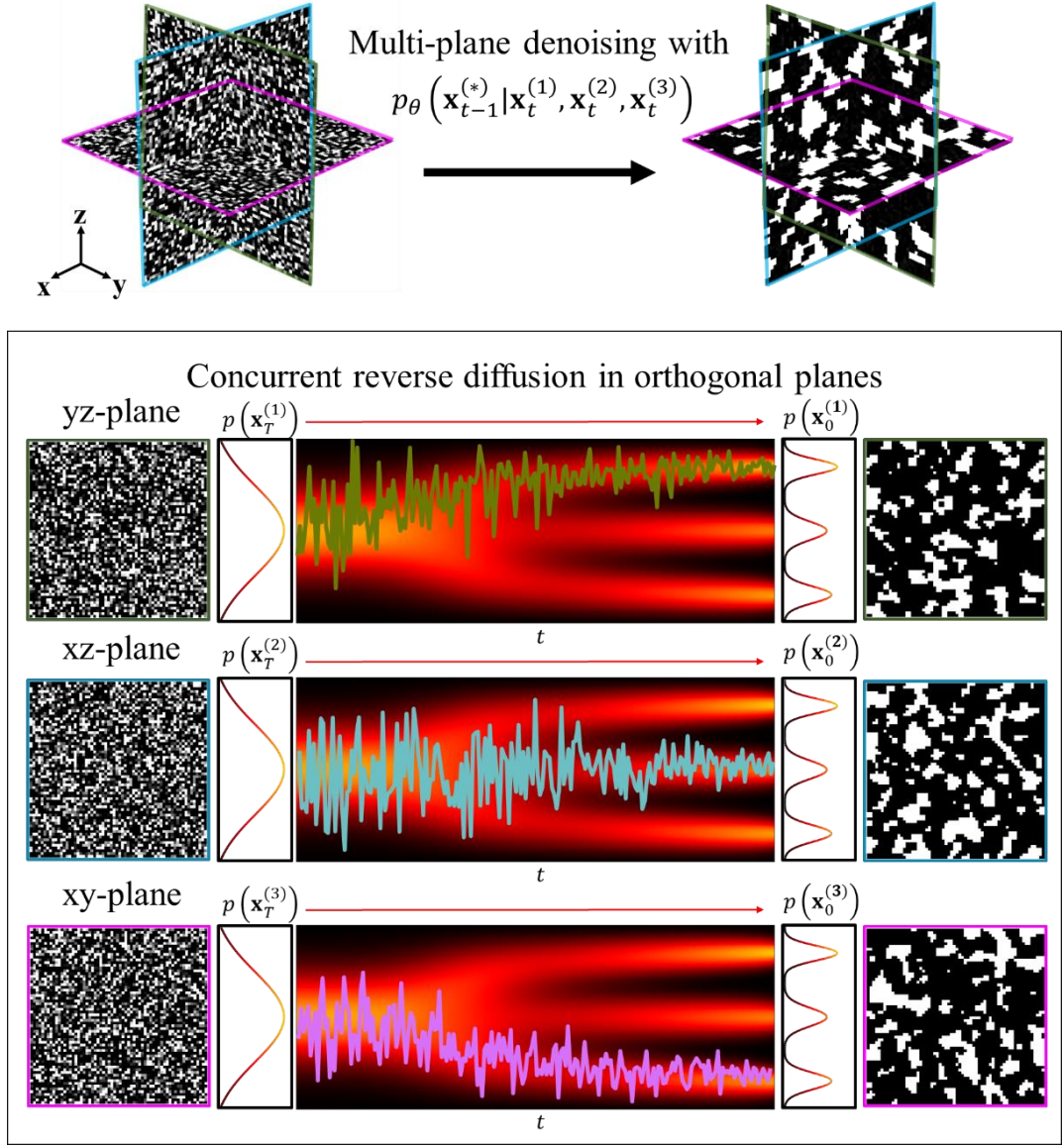

**Figure 2.** Schematic of multi-plane denoising diffusion in the three orthogonal planes (i.e., yz, xz, xy-planes) for transforming the noise distribution into the data distribution while ensuring connectivity between the planes. The colored lines within the time interval represent the trajectories of reverse diffusion ($p(\mathbf{x}_T) \to p(\mathbf{x}_0)$).

To avoid the preparation of a 3D dataset and maintain the architecture of 2D-DGM, the proposed method for 2D-to-3D dimensionality expansion in this study solely focuses on the reverse diffusion process. Consequently, the training process for the 2D-DGM remains unchanged, following the procedures used to train generative models for conventional image generation tasks with 2D datasets. Therefore, throughout this paper, the only neural network model we deal with is $p_\theta(\mathbf{x}_{t-1}|\mathbf{x}_t)$ trained on 2D microstructure images without any additional conditions incorporated. In this regard, the key idea of the proposed multi-plane denoising diffusion is to generate 3D voxels using 2D-DGM by changing the target denoising plane periodically as shown in Figure 3. The procedure of the multi-plane denoising diffusion is as follows:

1. Initialize voxels with dimensions of $n \times n \times n$ using Gaussian noise (Figure 3a). It is worth noting that the size of $n$ should match the size of input/output size of 2D-DGM (i.e., $n \times n$).

2. For the first reverse diffusion step, denoise the $n$ samples with dimensions of $n \times n$ (which are from the initialized voxels) using the trained 2D-DGM (i.e., $p_\theta(\mathbf{x}_{T-1}|\mathbf{x}_T)$) along a particular plane.

3. Before the second reverse diffusion step, change the target denoising plane and obtain $\mathbf{x}^*_{T-1}$, which represents the samples arranged in a different plane (Figure 3c). Then, denoise the samples by taking $\mathbf{x}^*_{T-1}$ as the input of the trained 2D-DGM (i.e., $p_\theta(\mathbf{x}^*_{T-2}|\mathbf{x}^*_{T-1})$). It is worth noting that $p_\theta(\mathbf{x}^*_{T-2}|\mathbf{x}^*_{T-1})$ is not a new model; it is the same as $p_\theta(\mathbf{x}_{T-1}|\mathbf{x}_T)$, but the input comes from a different plane.

4. Repeat the above process until the last reverse diffusion step is completed to obtain the 3D sample where all the images at different planes follow the data distribution $p(\mathbf{x}_0)$.

It is important to mention the proposed multi-plane denoising is based on the assumption that $\mathbf{x}^*_t$ closely resembles $\mathbf{x}_t$ since both the forward and reverse diffusion processes comprise multiple time steps to align the discrete DGMs with the continuous diffusion process in SDE formulation (section 2.1). In other words, the reverse Markovian process with the model $p_\theta(\mathbf{x}_{T-1}|\mathbf{x}_T)$ would function effectively with the input $\mathbf{x}^*_t$ because the denoising process operates in a gradual manner. Furthermore, since the denoising diffusion is performed at multiple planes together in a single reverse diffusion process, it allows for enforcing connectivity among the samples along different planes. Another perspective to comprehend this process is to consider $\mathbf{x}^*_t$ as the input of the model, which is manipulated before denoising using a transformation function $\psi_T$ (Figure 4). For instance, $\psi_T$ of the proposed multi-plane denoising is a function for rearranging $\mathbf{x}_t$ (i.e., manipulating rows and columns of samples to get $\mathbf{x}^*_t$) to change the target denoising plane. The interesting part is that $\psi_T$ can be modified for different purposes, such as rearranging the $\mathbf{x}_t$ to perform denoising diffusion at a much higher dimension and denoising multiple samples connected to each other for obtain a single high-resolution image (which is recommended for the future research works).

    However, one remaining issue is that $\mathbf{x}^*_t$ is not exactly same as $\mathbf{x}_t$ even with thousands of diffusion steps, although they may closely resemble each other. Due to the discrepancies in the distributions of $\mathbf{x}^*_t$ and $\mathbf{x}_t$, the Markov chain of $p_\theta(\mathbf{x}^*_{t-1}|\mathbf{x}^*_t)$ would produce lower-quality samples compared to those from $p_\theta(\mathbf{x}_{t-1}|\mathbf{x}_t)$. To fill the gap between $p_\theta(\mathbf{x}^*_{t-1}|\mathbf{x}^*_t)$ and $p_\theta(\mathbf{x}_{t-1}|\mathbf{x}_t)$ for generation of 3D sample with acceptable quality, the next section presents the method for harmonizing the samples at different planes.

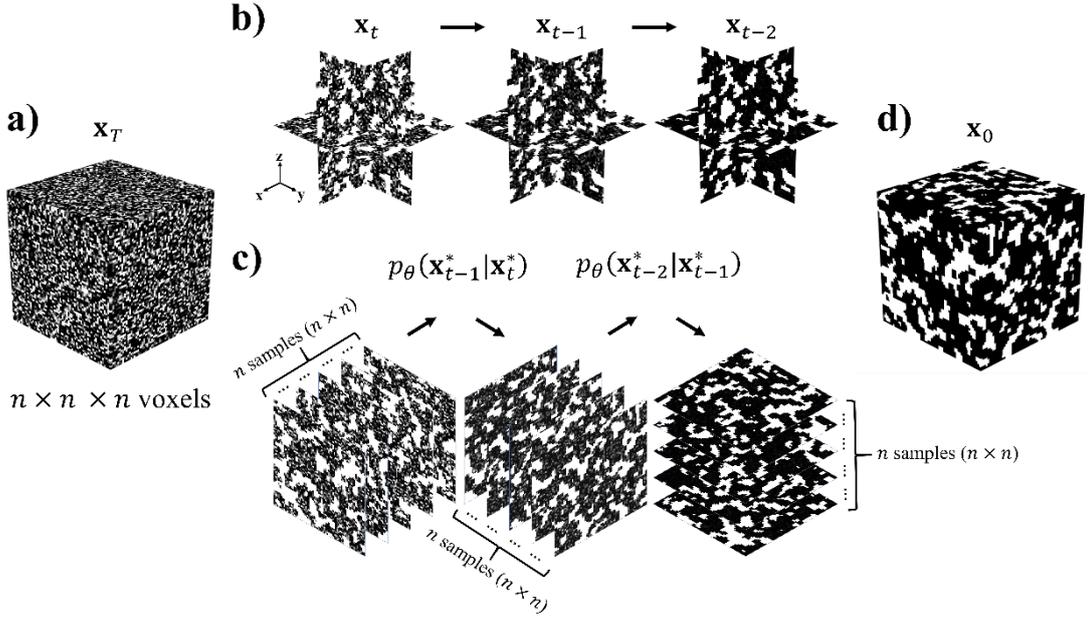

**Figure 3.** Multi-plane denoising with the discrete times steps of the reverse diffusion: a) Initial voxels with Gaussian noise, b) Slice view of the voxels according to the reversed time steps, c) Corresponding planes and samples for denoising at each time step with the trained model $p_\theta(\mathbf{x}_{t-1}|\mathbf{x}_t)$, and d) reconstructed 3D microstructure sample.

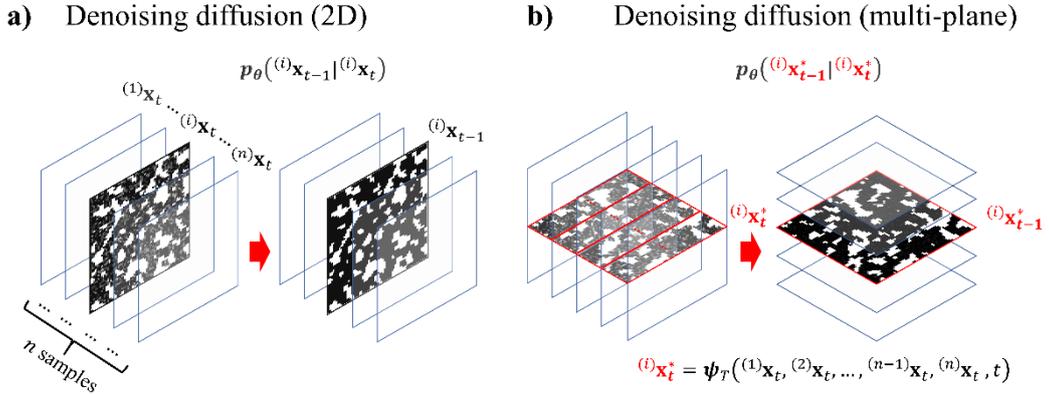

**Figure 4.** Comparison between a) conventional 2D denoising diffusion and b) multi-plane denoising diffusion in terms of the conditional probability and Markov chain based on the formulation of DDPM. The upper-left superscript denotes the *i*th sample along a particular plane.

### 3.3 Harmonized sampling for dimensionality expansion

To address the discrepancies caused by the proposed multi-plane denoising diffusion, this study introduces the method of harmonized sampling (or resampling [60]) for 2D-to-3D reconstruction of microstructures. Since $\mathbf{x}_t^*$ is manipulated data from $\mathbf{x}_t$, there is disharmony introduced to $\mathbf{x}_t^*$ which may lead to incorrect operations of $p_\theta(\mathbf{x}_{t-1}^*|\mathbf{x}_t^*)$. Although $p_\theta(\mathbf{x}_{t-1}^*|\mathbf{x}_t^*)$ would attempt to generate the most probable data with the estimation of $p(\mathbf{x}_{t-1}^*|\mathbf{x}_t^*)$ at every time step, the model cannot converge if it severely deviates from the correct trajectory at certain time step in the Markov chain of reverse diffusion (as the output at the current time step affects the output at the next time step). Therefore, a harmonizing step is adopted during the denoising process as depicted in Figure 5, which involves a cycle of renoising and denoising the sample at time $t$. In other words, we give more chances to the model to harmonize the conditional information $\mathbf{x}_t^*$ before proceeding to the next denoising step. The renoising process is the same as the original forward process (Eq. (8)) which can be written as follows.

$$p(\mathbf{x}_t^*|\mathbf{x}_{t-1}^*) = N(\mathbf{x}_t^*; \sqrt{1-\beta_t}\mathbf{x}_{t-1}^*, \beta_t \mathbf{I}) \qquad (21)$$

In addition, the concept of renoising (or harmonizing) was initially introduced for 'inpainting' with masked inputs to obtain the most probable inpainting result conditioned on the unmasked region by Lugmayr et al. [60]. Their work demonstrated that incorporating several harmonizing steps can result in more harmonized images, compared to sampling without harmonizing steps Taking inspiration from this idea, this study utilizes harmonizing steps to mitigate the disharmony introduced during the 2D-to-3D reconstruction with the multi-plane denoising diffusion (section 3.2). To be more specific, the harmonizing steps are applied $n_h$ times at each time step along with $\psi_T$ to enable sampling conditional to the samples at different planes. This approach allows us to obtain more realistic 3D data, ensuring that all the samples viewed from the three orthogonal planes follow the original data distribution and maintain connectivity. The entire process of multi-plane denoising diffusion with harmonizing steps for 2D-to-3D reconstruction of microstructures is described in Algorithm 1.

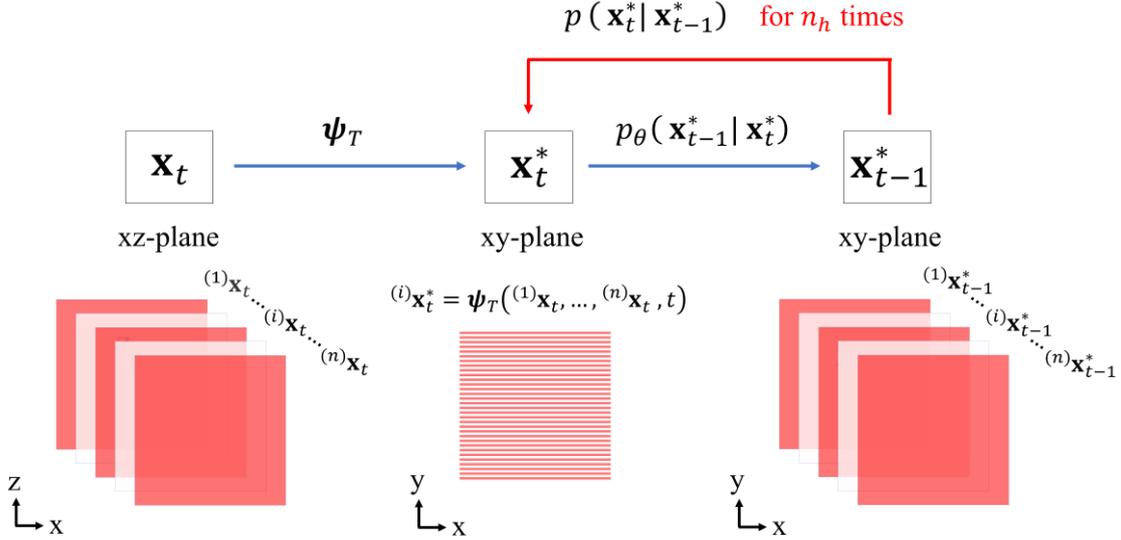

**Figure 5.** Multi-plane denoising diffusion with harmonizing steps to address the disharmony introduced by sampling in different planes.

---

**Algorithm 1:** Algorithm for multi-plane denoising diffusion for 2D-to-3D reconstruction of microstructure with harmonized sampling (the upper-left superscript for denoting the order of samples (Figure 4) is neglected for simplicity)

---

***Require:*** *the trained 2D diffusion model $\boldsymbol{\varepsilon}_\theta(\mathbf{x}_t, t)$ for generating $n \times n$ sized samples, the number of forward/reverse diffusion steps ($n_T$), the number of harmonizing step ($n_h$) (i.e., resampling step), the dimension of the volume to be generated ($n \times n \times n$ voxels), pre-defined noising schedule with $\beta_t$,* 1: *Create a class for re-noising a sample from $\mathbf{x}_{t-1}$ to $\mathbf{x}_t$ with the pre-defeind noising schedule i.e., $p(\mathbf{x}_t|\mathbf{x}_{t-1})$*
2: *Initialize $\mathbf{x}_T$ with a dimension of $n \times n \times n$ from a standard Gaussian distribution*
3: $\mathbf{x}_t^* = \mathbf{x}_T$
4: **for** $t = n_T, \ldots, 1$ **do**
5:     **for** $t_h = 1, \ldots, n_h$ **do**
6:         $\mathbf{z} \sim N(\mathbf{0}, \mathbf{I})$ if $t > 1$, else $\mathbf{z} = \mathbf{0}$
7:         $\mathbf{x}_{t-1}^* = \frac{1}{\sqrt{\alpha_t}}\left(\mathbf{x}_t^* - \frac{1-\alpha_t}{\sqrt{1-\bar{\alpha}_t}}\boldsymbol{\varepsilon}_\theta(\mathbf{x}_t^*, t)\right) + \sqrt{\beta_t}\mathbf{z}$
8:         $\mathbf{z}_h \sim N(\mathbf{0}, \mathbf{I})$
9:         Renoise ($\mathbf{x}_t^* = \sqrt{1-\beta_t}\mathbf{x}_{t-1} + \sqrt{\beta_t}\mathbf{z}_h$) if $t_h < n_h$ and $t < 1$
10:    **end for**
11:    Update $\mathbf{x}_{t-1}$ using $\mathbf{x}_{t-1}^*$: $\mathbf{x}_{t-1} = \boldsymbol{\psi}_{MP}^{-1}(\mathbf{x}_{t-1}^*, t)$
12:    Rearrange batch-wise input for multi-plane denoising diffusion: $\mathbf{x}_{t-1}^* = \boldsymbol{\psi}_T(\mathbf{x}_{t-1}, t-1)$
13: **end for**
14: **return** $\mathbf{x}_0$

---

### 3.4 Implementation

The implementation of the 2D-DGMs for sampling 2D microstructural images and the incorporation of the multi-plane denoising diffusion (section 3.2) with harmonizing steps (section 3.3) were carried out using the Pytorch library [61]. In order to build models for reverse diffusion process (i.e., $p_\theta(\mathbf{x}_{t-1}|\mathbf{x}_t)$), this study adopted the Imagen model, proposed by Saharia et al. [51] which proved to be highly effective in generating photorealistic images, to generate $64 \times 64$ sized 2D samples. The details of the model architecture with the hyperparameters are described in Appendix A.1.

The training of these models was conducted using Nvidia RTX A6000 graphics processing units (GPUs) coupled with the Adam optimizer, employing a learning rate of $3 \times 10^{-4}$. For each exemplary case, which will be further elaborated in the subsequent sections, the batch size was set to 32 per GPU, and the training process consisted of 50,000 training steps. The diffusion time steps for all models were set to be $T = 1000$ with the linear noise schedule [43], where $\beta_t$s are evenly spaced values over the interval $\beta_1 = 10^{-4}$ and $\beta_T = 10^{-2}$.

### 3.5 Evaluation metrics for validation

To quantify the quality of the generated samples using the proposed methodology in this study, the following criterion metrics are considered: two-point correlation function ($S_2$) and lineal path function ($LP$) in 2D/3D space. The two-point correlation function, which characterizes the statistical distribution of the material phase, can be written as follows:

$$S_2(\mathbf{r}_1, \mathbf{r}_2) = B(\mathbf{r}_1)B(\mathbf{r}_2) \tag{22}$$

where $B(\cdot)$ is a binary function that becomes 1 if a material phase of interest is present at a given location and 0 otherwise. In a similar manner, the lineal path function to evaluate the connectivity between clusters of material phase can be computed as follows:

$$LP(\mathbf{r}_1, \mathbf{r}_2) = \begin{cases} 1 & \text{if a line connecting } \mathbf{r}_1 \text{ and } \mathbf{r}_2 \text{ is on material phase of interest} \\ 0 & \text{otherwise} \end{cases} \quad (23)$$

Then, the error rate(%) between the correlation functions of training dataset and generated data can be computed using the following discrepancy equation [47, 62]:

$$Error\ rate\ (\%) = \frac{A_{dis}}{A_{ori}} \times 100 \quad (24)$$

where $A_{dis}$ represents the area between of two correlation functions and $A_{ori}$ represents the area under the correlation function of the training dataset. Depending on the sources of training data (e.g., 2D images sampled from 3D volume, or 2D images themselves), the correlation functions are evaluated in either 3D space or 2D space for each case study, which is further explained in the following sections

## 4. Case I: Synthetic microstructural samples of spherical inclusions

To validate the proposed multi-plane denoising diffusion in this study, 2D-to-3D reconstruction of microstructures with spherical inclusions embedded in a matrix was conducted. As shown in Figure 6, the 2D microstructure images used for training the 2D-DGM were sampled from a synthesized 3D data. The 3D sample of spherical inclusions was generated using Geodict® software by Math2Market GmbH. The diameters of the inclusions were set to follow a normal distribution with a mean value of 15 voxels and a standard deviation of 4 voxels in a volume of 64×64×300 voxels. Then, the 3D data was sliced along the z-axis, generating a total of 300 images with dimensions of $64 \times 64$. To facilitate the training of the 2D-DGM, the dataset was augmented by manipulating each image eight times (i.e., eight folds). This augmentation involves applying both horizontal and vertical flips, as well as rotations of 90°, 180°, and 270° to the original images.

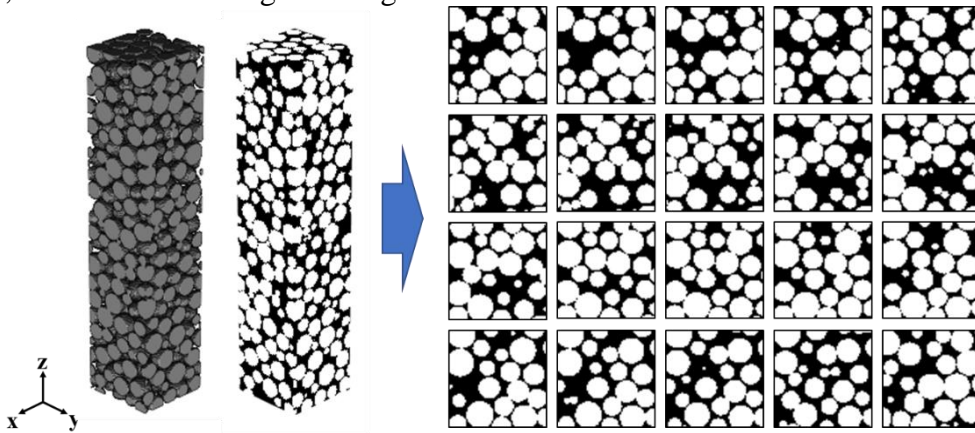

**Figure 6.** A synthesized 3D Microstructure sample with spherical inclusions and sampled 2D microstructural images for training the 2D-DGM.

After training the 2D-DGM, the multi-plane denoising with the harmonizing steps was conducted to create a 3D volume with spherical inclusions. To investigate the effect of number of harmonizing steps $n_h$, 2D-to-3D reconstruction was conducted with varying $n_h$ as illustrated in Figure 7. As depicted in the figure, the quality of the generated samples improves with an increase in the number of harmonizing steps. The results also show that the spherical shapes of the inclusions were well reproduced with $n_h = 10$. This implies that there is a minimal value of $n_h$ to guarantee the quality of the generated sample. For the sake of simplicity in the validation process, $n_h$ was set to be constant value of 10 for the example cases demonstrated in the following sections. Furthermore, the 3D visualization and sectional views in Figure 7 demonstrate that the inclusions (i.e., white pixels) along the three orthogonal planes are well connected, preserving continuity in 3D space.

To quantitatively evaluate the quality of the samples, the error rate (Eq. (24)) between the correlation functions of the original data (Figure 6) and generated data was evaluated, as shown in Figure 8a. For validating purpose, a total of 25 samples were generated for each case study using Micro3Diff to calculate the averaged $S_2$ and $LP$ (more examples of generated 3D samples can be found in Appendix A.2). In the case of spherical inclusions (case I), the error rates for $S_2$ and $L_p$ are 1.88% and 4.56% (Figure 9), respectively, indicating that the generated samples are in good agreement with the original data. In other words, the original and generated samples exhibit similar spatial correlations, indicating that they have statistically equivalent morphologies. Additionally, it is worth noting that since the source of training data (i.e., 2D images) for case I was sampled from the synthesized 3D volume, the 3D correlation functions were computed to evaluate the sample quality.

The results show that the multi-plane denoising diffusion with $p_\theta(\mathbf{x}^*_{t-1}|\mathbf{x}^*_t)$, aided by the harmonized sampling, effectively guides the distribution of images at different planes simultaneously to $p(\mathbf{x}_0)$, which is the distribution for images of spherical inclusions. It can also be said that the generation of 3D data with conditional distribution of images at different planes (Eq. (20)) is addressed by the dimensionality expansion problem using the gradual multi-plane diffusion process and harmonized sampling. In the following sections, additional example cases to evaluate the performance of Micro3Diff are explored, including the cases with the microstructure of polycrystalline grains with grain boundaries (case II) and the real-world experimental micrographs (case III and IV)

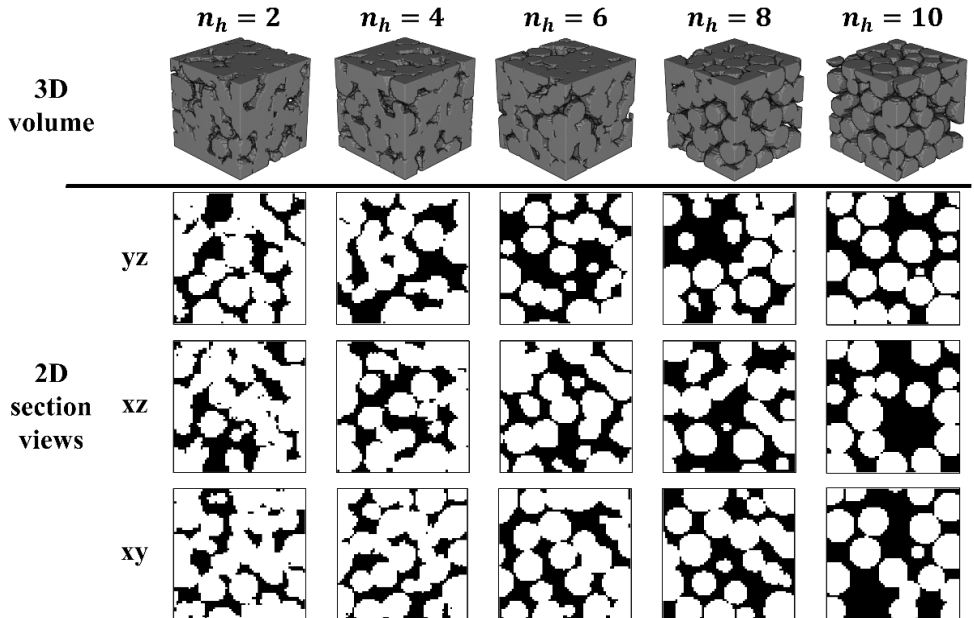

**Figure 7.** 2D-to-3D reconstruction results of microstructure with spherical inclusions using Micro3Diff with different numbers of harmonizing steps.

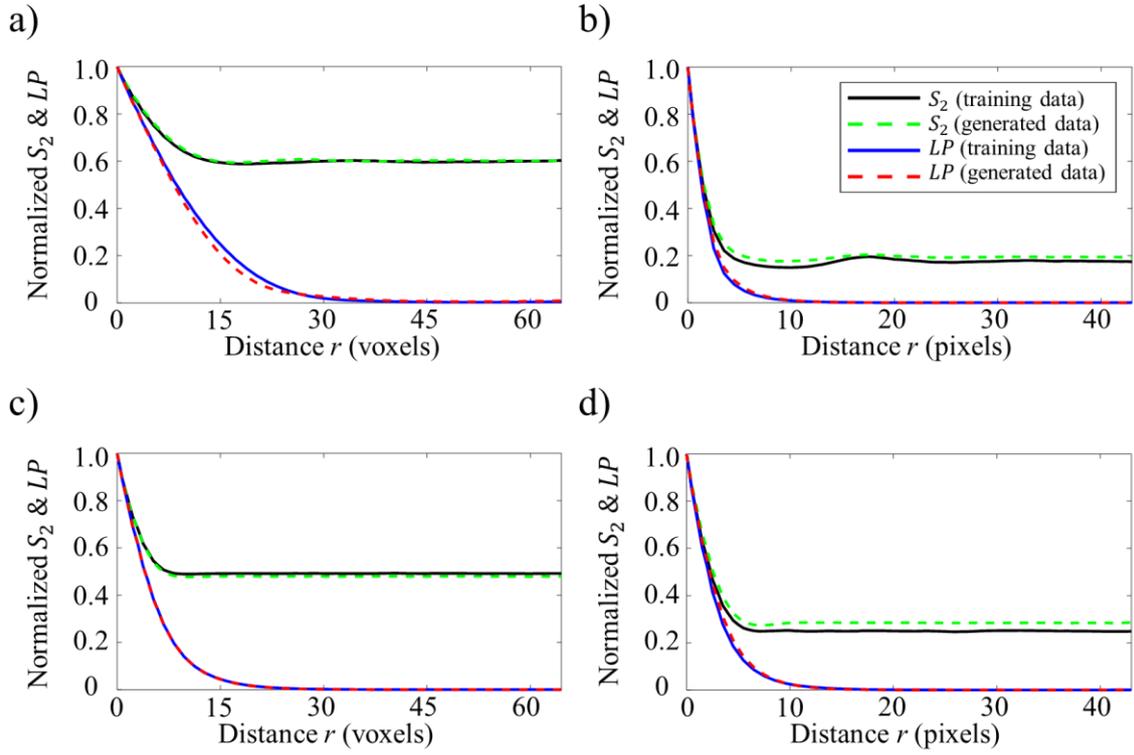

**Figure 8.** Two-point correlation functions ($S_2$) and lineal path functions ($LP$) for different types of microstructures: a) case I: spherical inclusions, b) case II: polycrystalline grains, c) case III: NMC battery cathode, and d): case IV: carbonates.

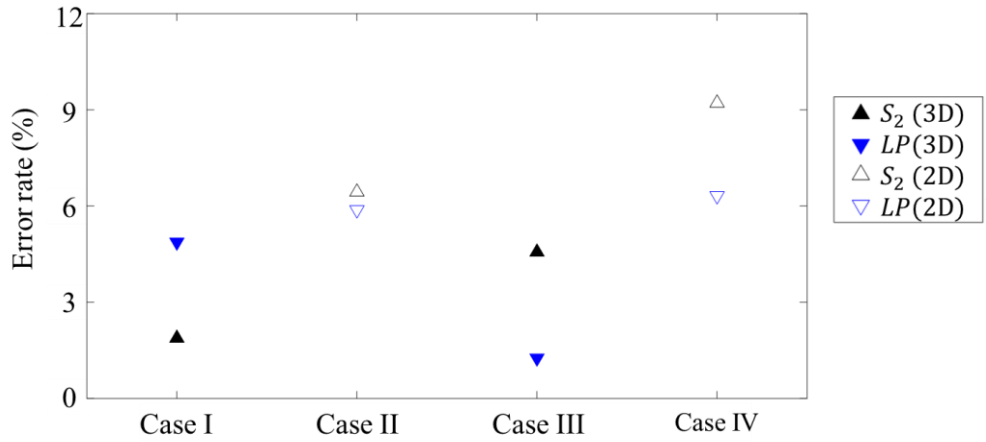

**Figure 9.** Error rate of two-point correlation function ($S_2$) and lineal path function ($L_p$) for different cases. For the cases where 2D samples from 3D data (i.e., 3D micro-CT data) were used for training the 2D-DGM (Case I and Case III), the error rate is evaluated in 3D space. For the cases where 2D samples were used for training the 2D-DGM (Case II and Case IV), the error rate is evaluated in 2D space.

## 5. Case II: Synthetic microstructures of polycrystalline grains

For further assessing the effectiveness of proposed Micro3Diff, this section focuses on the microstructural images of polycrystalline grains as the case study. To acquire 2D images of polycrystalline grains for the 2D-to-3D reconstruction task, a set of 300 synthetic images were generated using Laguerre tessellation [22] as shown in Figure 9. In each generated 2D image, the white pixels represent the grain boundaries. Three values of grain sizes were considered (6, 9 and 12 pixels) to be distributed, each occupying an equal fraction within a $64 \times 64$ image. The synthesized images were then augmented eightfold, similar to case I for training the 2D-DGM.

It is worth noting that the synthetic 2D samples of polycrystalline grains (Figure 10) considered in this case are genuinely independent samples, meaning that they were not sampled from a 3D data as in case I (section 4). Consequently, while the 2D-DGM lacks the ability to learn the 3D relationship for both cases I and II, the authors hypothesized that case II poses a more challenging problem since it lacks an intermediate image that can connect different slices. Thus, the Markov chain of $p_\theta(\mathbf{x}_{t-1}^*|\mathbf{x}_t^*)$ may have a higher likelihood of deviating from the trajectories of the Markov chain of $p_\theta(\mathbf{x}_{t-1}|\mathbf{x}_t)$.

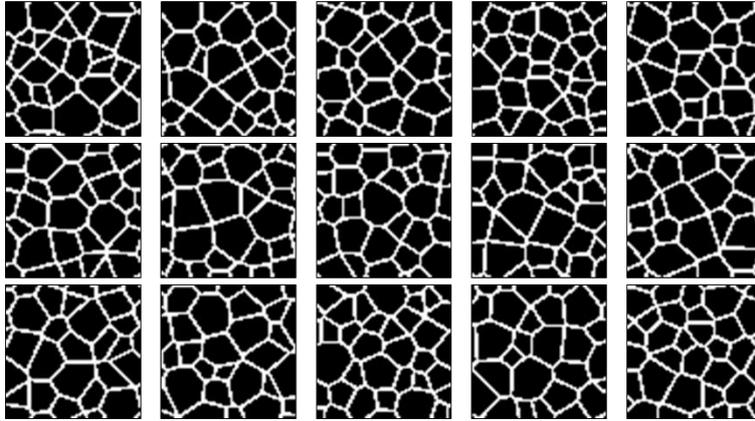

**Figure 10.** Synthesized 2D microstructure samples of polycrystalline grains for training the 2D-DGM.

Figure 10 shows a generated 3D volume with polycrystalline grains using Micro3Diff, along with its sectional views at each orthogonal plane. To show the connectivity and the grain structures clearly, the 3D volumetric view includes the normal view, where only the grain boundaries (i.e., white voxels) are visualized, and its inverted view. At first glance, the grain boundaries appear to be well connected in 3D space, which is the important feature of polycrystalline materials. However, the 2D sectional views show that there are some disconnected and curved grain boundaries in the generated 3D volume. This error is likely due to the lack of intermediate images in the training dataset for connecting different slices in the 3D volume, which makes the model to generate sample with lower probabilities in $p(\mathbf{x}_0)$. In addition, the occurrence of disconnected grain boundaries has been commonly reported even in the recent studies with the descriptor-based MCR method [17, 19] and GAN-based reconstruction [34]. Meanwhile, the error rates for $S_2$ and $LP$ are 6.43% and 5.88%, respectively, indicating the good agreement between the original and generated data (Figure 8b and Figure 9). It is worth noting that, since the source of the training data for case II is 2D images (Figure 10), the 2D correlation functions were computed to evaluate the generated sample quality at the three orthogonal planes. Then, the mean values of the computed correlation functions were used to evaluate the error rate. To enhance the performance of 2D-to-3D reconstruction for microstructures of polycrystal materials such as alloys, the authors suggest considering preprocessing and augmentation of dataset to address the lack of intermediate images, as well as optimizing the hyperparameters (e.g., $n_h$ and $T$)

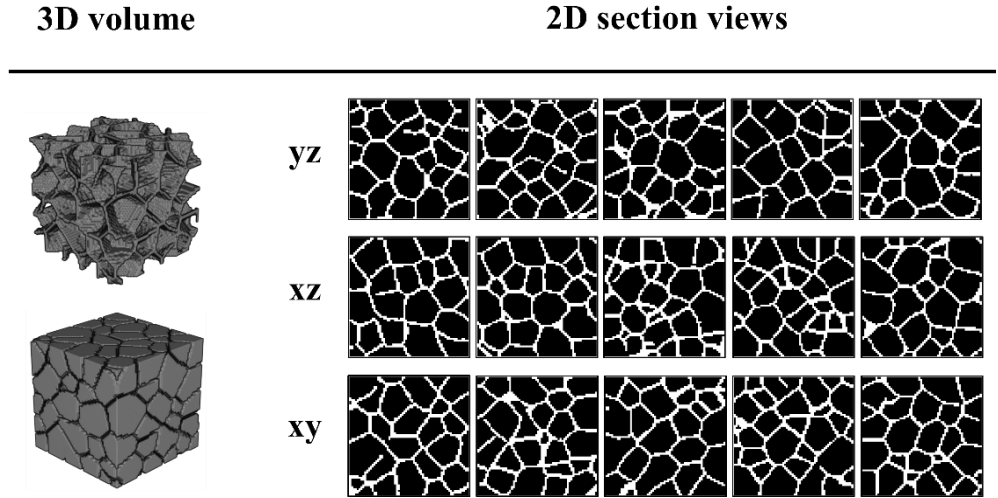

**Figure 11.** 2D-to-3D reconstruction results of microstructure with polycrystalline grains using Micro3Diff.

## 6. Case III: Experimental micro-CT scan images of battery electrodes

The next example case is the real micro-CT scan images of nickel manganese cobalt (NMC) cathode. The micro-CT data of NMC cathode was obtained from open-access collected data in reference [63], and a volume of 64×64×300 voxels was created for sampling 300 images with dimensions of $64 \times 64$ (Figure 12). Subsequently, the sampled images were augmented using eightfold augmentation as the previous cases.

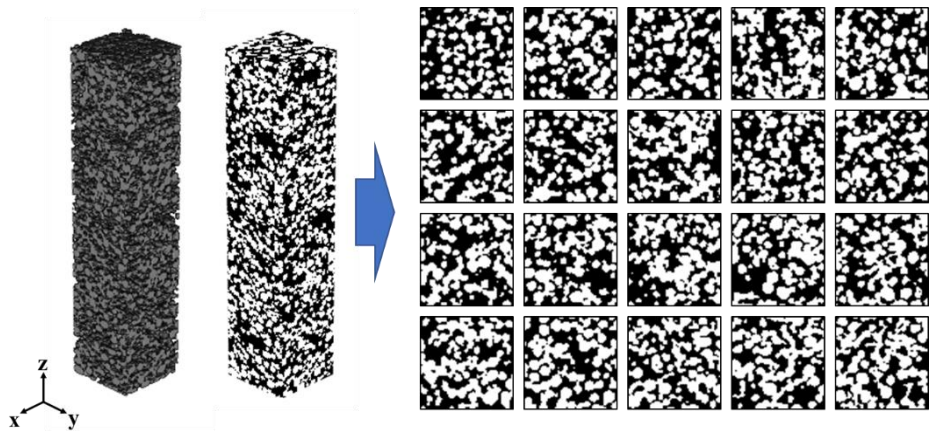

**Figure 12.** An experimentally obtained 3D microstructure sample of NMC battery cathode and sampled 2D microstructural images for training the 2D-DGM.

Figure 13 illustrates the generated 3D volume of NMC cathode and its 2D sectional views (for simplicity, only the active material phase and pore were considered). Similar to case I and II, the results show that the connectivity of the white voxels (i.e., active material phase) is preserved. Both the 3D volume and the 2D sectional views exhibit visual similarity to the original data in Figure 12. The error rates of 4.56% for $S_2$ and 1.26% for $LP$ (Figure 8c and Figure 9) also demonstrate that the generated samples have close spatial correlations compared to the original micro-CT data. This is particularly encouraging as it demonstrates the effectiveness of the proposed Micro3Diff in 2D-to-3D reconstruction, utilizing both synthetic (case I and II) and experimentally observed data. Since obtaining a reliable dataset of microstructure is crucial for characterizing the material behavior, there has been significant attention given to the reconstruction of battery electrode microstructures in recent years [64-66]. Although the performance of the proposed methodology must be validated with more diverse experimental data, the results highlight the potential application of Micro3Diff in the systematic exploration and design of materials for battery applications aided by ICME methods [3, 4].

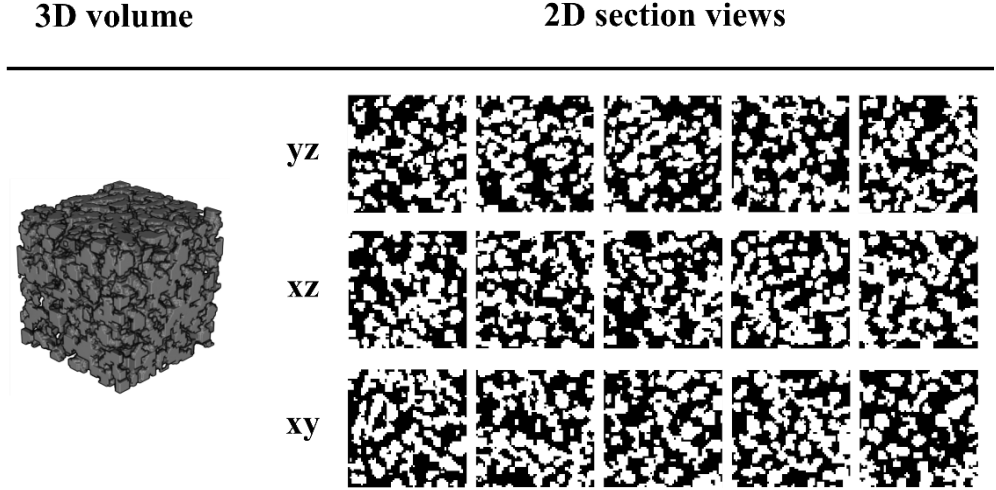

**Figure 13.** 2D-to-3D reconstruction results for microstructure of NMC battery cathode using Micro3Diff.

## 7. Case IV: Experimental micro-CT scan image of carbonates

Lastly, a more challenging scenario is considered where the availability of 3D microstructure data is severely restricted, and only obtainable data is a single representative 2D micrograph image. Suppose a single 2D sample of carbonates [67] is available, and we aim to reconstruct a 3D volume from this 2D sample, which should have an equivalent statistical distribution of material phases in 3D space. To train 2D-DGM first, multiple cropped images can be sampled from the representative image as shown in Figure 14. In particular, a total of 250 images were sampled and augmented using eightfold augmentation, as in the previous cases.

Figure 15 shows the generated 3D volume of carbonates with its sectional views. As can be seen, the generated 3D volume exhibits visually similar sectional views at the three orthogonal planes, demonstrating the capability of Micro3Diff to create a 3D volume from a single experimental representative image. Meanwhile, the maximum error rates for $S_2$ and *LP* are 9.21% and 6.31%, respectively (Figure 8d and Figure 9). The higher error rates compared to the previous cases are likely due to the lack of diversity in the data used to train the 2D-DGM, which could lead to incorrect estimation of the data distribution during the multi-plane denoising process. In other words, the absence of intermediate data to ensure connectivity in 3D space could be a reason for the higher error rates (similar to case II). However, the results are remarkably encouraging, as the sectional views display morphological similarities with the training data, and the volumetric view illustrates Micro3Diff's ability to discover potential combinations of 2D images from the estimated $p(\mathbf{x}_0)$ (derived from a single micrograph) to construct a 3D volume. Moreover, the proposed Micro3Diff enables the generation of multiple and diverse volumes with comparable visuals and spatial correlations to the original data, as shown in Appendix A.2. This capability could significantly assist in the quantitative assessment of material behavior in 3D space [13, 68-70], taking into account the microstructures and inherent randomness.

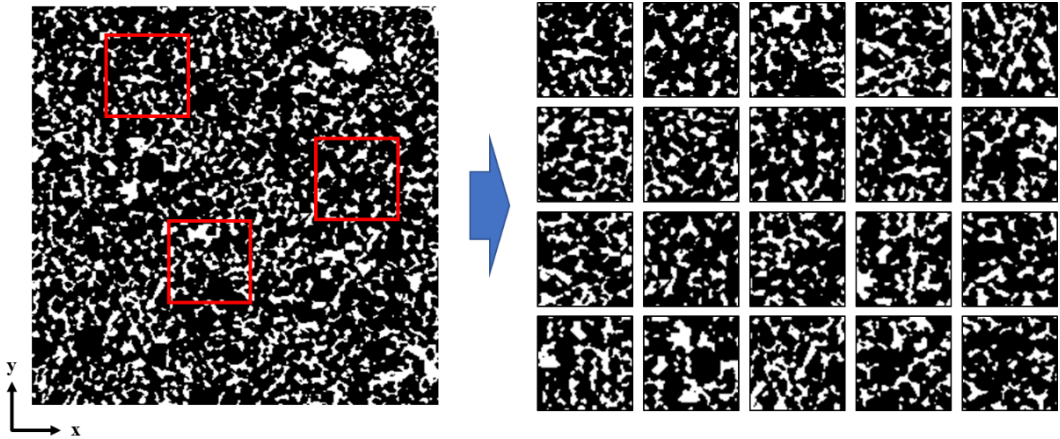

**Figure 14.** An experimentally obtained single 2D microstructure image of carbonates and random sampling of 2D images for training the 2D-DGM.

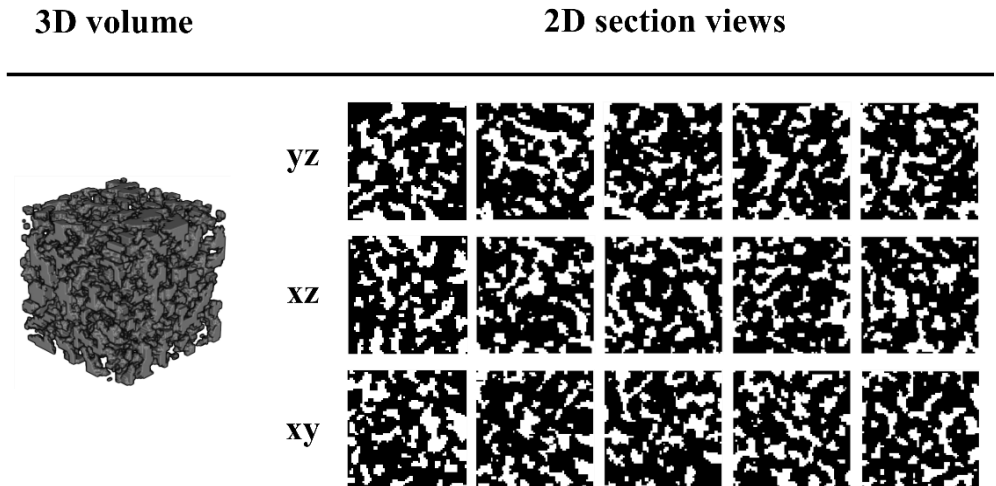

**Figure 15.** 2D-to-3D reconstruction results for microstructure of carbonates using Micro3Diff.

## 8. Conclusions

The results in this study demonstrate the capability of Micro3Diff for reconstructing 3D microstructure samples from 2D micrographs using the proposed multi-plane denoising diffusion and the harmonized sampling. Based on the multi-plane denoising, a 2D-DGM can be used to effectively guide the noised samples across multiple planes towards the data distribution while maintaining connectivity in 3D space. The disharmony caused by potential deviations from the intended trajectories in the trained 2D noise-to-data denoising diffusion process can be addressed by the proposed harmonized sampling method. Since this is the first study that proposes DGM-based 2D-to-3D microstructure reconstruction, the authors believe that the proposed Micro3Diff has the potential to chart novel avenues in the realm of data-driven material science. Meanwhile, a broader range of microstructure types, including multi-phase and anisotropic microstructures, should be considered to improve the applicability of Micro3Diff in the future. In addition, the proposed Micro3Diff also introduces a novel approach of manipulating the latent variables (i.e., noised samples) in DGMs for dimensionality expansion. The authors suggest that this concept could substantially expand the range of applications for DGMs in various fields, including computational materials engineering and 3D generative modeling.

**Acknowledgments**

This material is based upon work supported by the Air Force Office of Scientific Research under award number FA2386-22-1-4001 and the Institute of Engineering Research at Seoul National University.

**Data availability**
The study used open-access microstructural data for training models from the following sources: battery electrodes (NMC cathodes) [63] and carbonates [67]. All generated data used are available from the authors on request.

**Competing interests**
The authors declare no competing interests.

**Appendix**
A.1 Parameters of 2D-DGM

In this study, the 2D-DGM is built using the architecture of Imagen proposed by Saharia et al. [51]. In addition, only the $64 \times 64$ base model (which utilizes the U-Net architecture [71]) is used for implementation, without incorporating the super-resolution cascade for sampling higher-resolution images as proposed in the original paper. The specific parameters used for the 2D-DGM in this study are presented in Table A 1.

**Table A 1.** Parameters of 2D-DGMs used for 2D-to-3D reconstruction of microstructures (i.e., Micro3Diff) in this study

| DGM parameters | |
| --- | --- |
| Image size | $\mathbf{64 \times 64}$ |
| Number of forward diffusion (i.e., noising) steps | **1000** |
| Noising schedule | Linear |
| Optimizer | Adam optimizer; learning rate of $\mathbf{3 \times 10^{-4}}$; **10000** warm-up steps |
| Batch size | 64 |
| Hyperparameters of U-Net (in DGMs) | |
| Dimension | **128** |
| Dimension multipliers | $\mathbf{1, 2, 4, 8}$ |
| Layer attentions | False, False, True, True |
| Layer cross-attentions | False |
| Number of Resnet blocks | **2** |
| Number of attention heads | **8** |
| Dimension of attention head | **64** |

A.2 Generated 3D samples with Micro3Diff

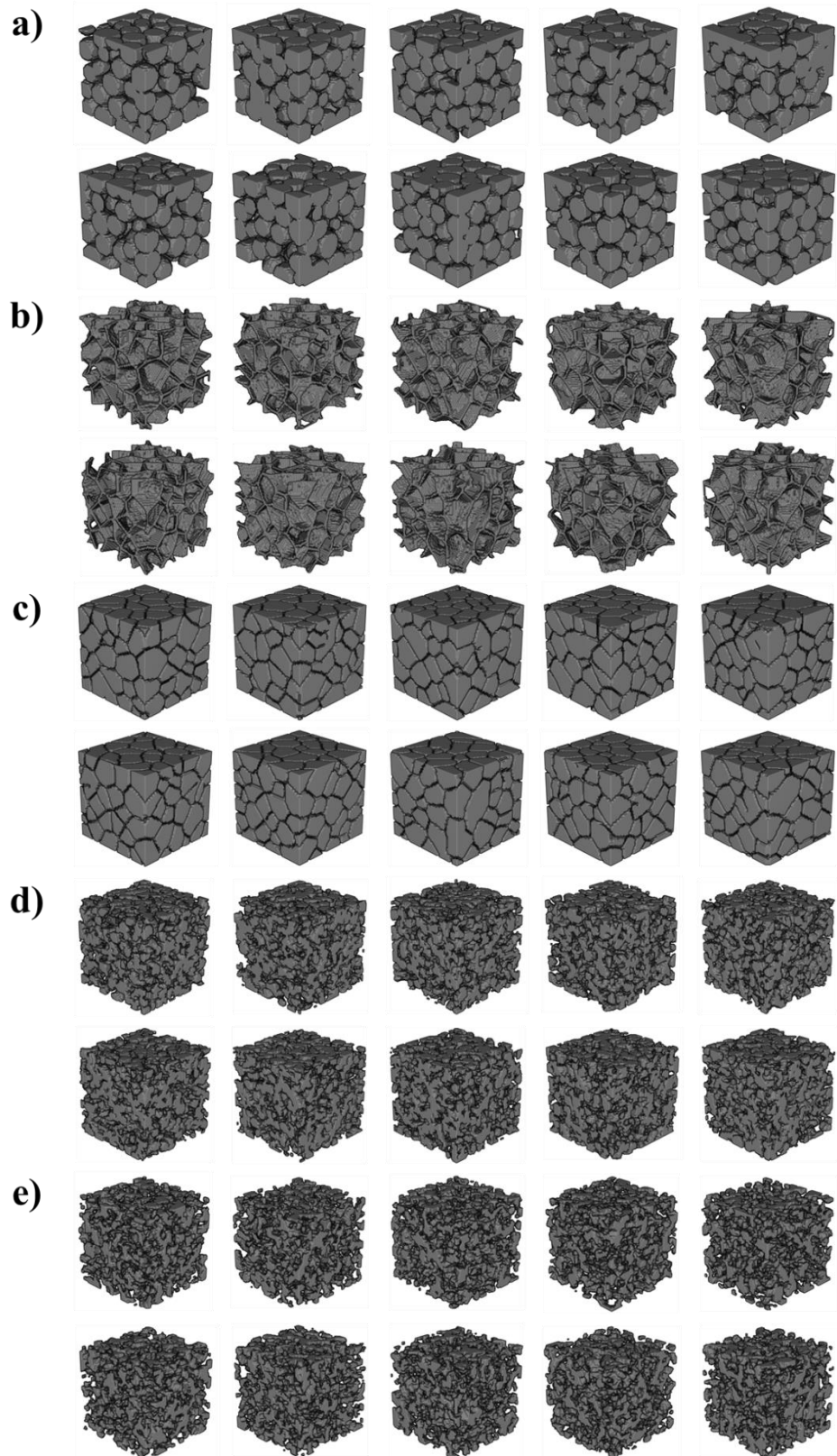

**Figure A 1.** Exemplary 2D-to-3D reconstruction results of different microstructures using Micro3Diff: a) spherical inclusions, b) polycrystalline grains, c) polycrystalline grains (inverted), d) NMC battery electrode, and d) carbonate

**References**
[1] Ghosh S, Dimiduk D. Computational methods for microstructure-property relationships: Springer; 2011.
[2] Fish J, Wagner GJ, Keten S. Mesoscopic and multiscale modelling in materials. Nature materials. 2021;20(6):774-86.
[3] Allison J. Integrated computational materials engineering: A perspective on progress and future steps. Jom. 2011;63(4):15.
[4] Allison J, Backman D, Christodoulou L. Integrated computational materials engineering: a new paradigm for the global materials profession. Jom. 2006;58:25-7.
[5] Lee K-H, Lim HJ, Yun GJ. A Data-Driven Framework for Designing Microstructure of Multifunctional Composites with Deep-Learned Diffusion-Based Generative Models. 2023.
[6] Vlassis NN, Sun W. Denoising diffusion algorithm for inverse design of microstructures with fine-tuned nonlinear material properties. Computer Methods in Applied Mechanics Engineering. 2023;413:116126.
[7] Lee XY, Waite JR, Yang C-H, Pokuri BSS, Joshi A, Balu A, et al. Fast inverse design of microstructures via generative invariance networks. Nature Computational Science. 2021;1(3):229-38.
[8] Horstemeyer MF. Integrated Computational Materials Engineering (ICME) for metals: using multiscale modeling to invigorate engineering design with science: John Wiley & Sons; 2012.
[9] Bargmann S, Klusemann B, Markmann J, Schnabel JE, Schneider K, Soyarslan C, et al. Generation of 3D representative volume elements for heterogeneous materials: A review. Progress in Materials Science. 2018;96:322-84.
[10] Maire E, Buffiere J-Y, Salvo L, Blandin JJ, Ludwig W, Letang JM. On the Application of X-ray Microtomography in the Field of Materials Science. Advanced Engineering Materials. 2001;3(8):539-46.
[11] Lee K-H, Lee HW, Yun GJ. A defect detection framework using three-dimensional convolutional neural network (3D-CNN) with in-situ monitoring data in laser powder bed fusion process. Optics Laser Technology. 2023;165:109571.
[12] Lim HJ, Choi H, Lee MJ, Yun GJ. An efficient multi-scale model for needle-punched Cf/SiCm composite materials with experimental validation. Composites Part B: Engineering. 2021;217:108890.
[13] Bostanabad R, Zhang Y, Li X, Kearney T, Brinson LC, Apley DW, et al. Computational microstructure characterization and reconstruction: Review of the state-of-the-art techniques. Progress in Materials Science. 2018;95:1-41.
[14] Geers MG, Kouznetsova VG, Brekelmans W. Multi-scale computational homogenization: Trends and challenges. Journal of computational applied mathematics. 2010;234(7):2175-82.
[15] Rao C, Liu Y. Three-dimensional convolutional neural network (3D-CNN) for heterogeneous material homogenization. Computational Materials Science. 2020;184:109850.
[16] Yvonnet J. Computational homogenization of heterogeneous materials with finite elements: Springer; 2019.
[17] Seibert P, Raßloff A, Ambati M, Kästner M. Descriptor-based reconstruction of three-dimensional microstructures through gradient-based optimization. Acta Materialia. 2022;227:117667.
[18] Xu H, Dikin DA, Burkhart C, Chen W. Descriptor-based methodology for statistical characterization and 3D reconstruction of microstructural materials. Computational Materials Science. 2014;85:206-16.
[19] Seibert P, Raßloff A, Kalina K, Ambati M, Kästner M. Microstructure Characterization and Reconstruction in Python: MCRpy. Integrating Materials Manufacturing Innovation. 2022;11(3):450-66.
[20] Seibert P, Raßloff A, Kalina KA, Gussone J, Bugelnig K, Diehl M, et al. Two-stage 2D-to-3D reconstruction of realistic microstructures: Implementation and numerical validation by effective properties. Computer Methods in Applied Mechanics Engineering optimization. 2023;412:116098.
[21] Li K-Q, Liu Y, Yin Z-Y. An improved 3D microstructure reconstruction approach for porous media. Acta Materialia. 2023;242:118472.
[22] Torquato S. Statistical description of microstructures. Annual review of materials research. 2002;32(1):77-111.
[23] Torquato S, Haslach Jr H. Random heterogeneous materials: microstructure and macroscopic properties. Appl Mech Rev. 2002;55(4):B62-B3.
[24] Yeong C, Torquato S. Reconstructing random media. Physical review E. 1998;57(1):495.
[25] Jiao Y, Stillinger F, Torquato S. Modeling heterogeneous materials via two-point correlation functions: Basic principles. Physical review E. 2007;76(3):031110.
[26] Lu B, Torquato S. Lineal-path function for random heterogeneous materials. Physical Review A. 1992;45(2):922.
[27] Seibert P, Ambati M, Raßloff A, Kästner M. Reconstructing random heterogeneous media through differentiable optimization. Computational Materials Science. 2021;196:110455.
[28] Kim Y, Park HK, Jung J, Asghari-Rad P, Lee S, Kim JY, et al. Exploration of optimal microstructure and mechanical properties in continuous microstructure space using a variational autoencoder. Materials Design. 2021;202:109544.

[29] Sundar S, Sundararaghavan V. Database development and exploration of process–microstructure relationships using variational autoencoders. 2020;25:101201.
[30] Noguchi S, Inoue J. Stochastic characterization and reconstruction of material microstructures for establishment of process-structure-property linkage using the deep generative model. Physical Review E. 2021;104(2):025302.
[31] Xu L, Hoffman N, Wang Z, Xu H. Harnessing structural stochasticity in the computational discovery and design of microstructures. Materials Design. 2022;223:111223.
[32] Gayon-Lombardo A, Mosser L, Brandon NP, Cooper SJ. Pores for thought: generative adversarial networks for stochastic reconstruction of 3D multi-phase electrode microstructures with periodic boundaries. npj Computational Materials. 2020;6(1):1-11.
[33] Fokina D, Muravleva E, Ovchinnikov G, Oseledets I. Microstructure synthesis using style-based generative adversarial networks. Physical Review E. 2020;101(4):043308.
[34] Kench S, Cooper SJ. Generating three-dimensional structures from a two-dimensional slice with generative adversarial network-based dimensionality expansion. Nature Machine Intelligence. 2021;3(4):299-305.
[35] Kench S, Cooper SJ. Generating 3D structures from a 2D slice with GAN-based dimensionality expansion. arXiv preprint arXiv:07708. 2021.
[36] Zhang F, Teng Q, Chen H, He X, Dong X. Slice-to-voxel stochastic reconstructions on porous media with hybrid deep generative model. Computational Materials Science. 2021;186:110018.
[37] Tolstikhin I, Bousquet O, Gelly S, Schoelkopf B. Wasserstein auto-encoders. arXiv preprint arXiv:01558. 2017.
[38] Li Y, Swersky K, Zemel R. Generative moment matching networks.  International conference on machine learning: PMLR; 2015. p. 1718-27.
[39] Lala S, Shady M, Belyaeva A, Liu M. Evaluation of mode collapse in generative adversarial networks. High Performance Extreme Computing. 2018.
[40] Miyato T, Kataoka T, Koyama M, Yoshida Y. Spectral normalization for generative adversarial networks. arXiv preprint arXiv:05957. 2018.
[41] Song Y, Sohl-Dickstein J, Kingma DP, Kumar A, Ermon S, Poole B. Score-based generative modeling through stochastic differential equations. arXiv preprint arXiv:13456. 2020.
[42] Song Y, Ermon S. Generative modeling by estimating gradients of the data distribution. Advances in Neural Information Processing Systems. 2019;32.
[43] Ho J, Jain A, Abbeel P. Denoising diffusion probabilistic models. Advances in Neural Information Processing Systems. 2020;33:6840-51.
[44] Vincent PJNc. A connection between score matching and denoising autoencoders. 2011;23(7):1661-74.
[45] Yang L, Zhang Z, Song Y, Hong S, Xu R, Zhao Y, et al. Diffusion models: A comprehensive survey of methods and applications. arXiv preprint arXiv:00796. 2022.
[46] Dhariwal P, Nichol A. Diffusion models beat gans on image synthesis. Advances in Neural Information Processing Systems. 2021;34:8780-94.
[47] Lee K-H, Yun GJ. Microstructure reconstruction using diffusion-based generative models. Mechanics of Advanced Materials Structures. 2023:1-19.
[48] Düreth C, Seibert P, Rücker D, Handford S, Kästner M, Gude M. Conditional diffusion-based microstructure reconstruction. Materials Today Communications. 2023;35:105608.
[49] Song J, Meng C, Ermon S. Denoising diffusion implicit models. arXiv preprint arXiv:02502. 2020.
[50] Cao H, Tan C, Gao Z, Chen G, Heng P-A, Li SZ. A survey on generative diffusion model. arXiv preprint arXiv:02646. 2022.
[51] Saharia C, Chan W, Saxena S, Li L, Whang J, Denton E, et al. Photorealistic Text-to-Image Diffusion Models with Deep Language Understanding. arXiv preprint arXiv:11487. 2022.
[52] Rombach R, Blattmann A, Lorenz D, Esser P, Ommer B. High-resolution image synthesis with latent diffusion models.  Proceedings of the IEEE/CVF Conference on Computer Vision and Pattern Recognition2022. p. 10684-95.
[53] Song Y, Ermon S. Improved techniques for training score-based generative models. Advances in neural information processing systems. 2020;33:12438-48.
[54] Croitoru F-A, Hondru V, Ionescu RT, Shah M. Diffusion models in vision: A survey. IEEE Transactions on Pattern Analysis Machine Intelligence. 2023.
[55] Øksendal B, Øksendal B. Stochastic differential equations: Springer; 2003.
[56] Anderson BD. Reverse-time diffusion equation models. Stochastic Processes and their Applications. 1982;12(3):313-26.
[57] Kingma DP, Welling M. An introduction to variational autoencoders. Foundations Trends® in Machine Learning. 2019;12(4):307-92.

[58] Nichol AQ, Dhariwal P. Improved denoising diffusion probabilistic models.   International Conference on Machine Learning: PMLR; 2021. p. 8162-71.
[59] Ho J, Salimans T. Classifier-free diffusion guidance. arXiv preprint arXiv:12598. 2022.
[60] Lugmayr A, Danelljan M, Romero A, Yu F, Timofte R, Van Gool L. Repaint: Inpainting using denoising diffusion probabilistic models.   Proceedings of the IEEE/CVF Conference on Computer Vision and Pattern Recognition2022. p. 11461-71.
[61] Paszke A, Gross S, Massa F, Lerer A, Bradbury J, Chanan G, et al. Pytorch: An imperative style, high-performance deep learning library. Advances in neural information processing systems. 2019;32.
[62] Li X, Zhang Y, Zhao H, Burkhart C, Brinson LC, Chen W. A transfer learning approach for microstructure reconstruction and structure-property predictions. Scientific reports. 2018;8(1):1-13.
[63] Battery Microstructure Li-Ion Cathode and Anode Data Samples. National Renewable Energy Laboratory.
[64] Xu H, Zhu J, Finegan DP, Zhao H, Lu X, Li W, et al. Guiding the design of heterogeneous electrode microstructures for Li-ion batteries: microscopic imaging, predictive modeling, and machine learning. Advanced Energy Materials. 2021;11(19):2003908.
[65] Kim S, Wee J, Peters K, Huang H-YS. Multiphysics coupling in lithium-ion batteries with reconstructed porous microstructures. The Journal of Physical Chemistry C. 2018;122(10):5280-90.
[66] Lu X, Bertei A, Finegan DP, Tan C, Daemi SR, Weaving JS, et al. 3D microstructure design of lithium-ion battery electrodes assisted by X-ray nano-computed tomography and modelling. Nature communications. 2020;11(1):2079.
[67] Prodanovic M, Esteva M, Hanlon M, Nanda G, Agarwal P. Digital Rocks Portal: a repository for porous media images. 10.17612. P7CC7K; 2015.
[68] Li K-Q, Li D-Q, Liu Y. Meso-scale investigations on the effective thermal conductivity of multi-phase materials using the finite element method. International Journal of Heat Mass Transfer. 2020;151:119383.
[69] Rüger B, Joos J, Weber A, Carraro T, Ivers-Tiffée E. 3D electrode microstructure reconstruction and modelling. ECS Transactions. 2009;25(2):1211.
[70] Kumar H, Briant C, Curtin W. Using microstructure reconstruction to model mechanical behavior in complex microstructures. Mechanics of Materials. 2006;38(8-10):818-32.
[71] Ronneberger O, Fischer P, Brox T. U-net: Convolutional networks for biomedical image segmentation. International Conference on Medical image computing and computer-assisted intervention: Springer; 2015. p. 234-41.